\documentclass[10pt,conference]{IEEEtran}

\usepackage[T1]{fontenc}
\usepackage[utf8]{inputenc}
\usepackage{newtxmath}
\usepackage{csquotes}
\usepackage{xspace}
\usepackage{textcomp}
\usepackage{soul}
 
\usepackage{amsmath,amsfonts}
\usepackage{graphicx}

\usepackage{color}
\usepackage{xcolor}
\usepackage{algorithm}
\usepackage{algpseudocode}
\usepackage{listings}
\usepackage{balance}
\usepackage{hyperref}
\hypersetup{hidelinks}
\usepackage{tabularx}
\usepackage{array}
\usepackage{makecell}
\definecolor{codegreen}{rgb}{0,0.6,0}
\definecolor{codegray}{rgb}{0.5,0.5,0.5}
\definecolor{codepurple}{rgb}{0.58,0,0.82}
\definecolor{backcolour}{rgb}{0.95,0.95,0.92}
\algrenewcommand\algorithmicrequire{\textbf{Input:}}
\algrenewcommand\algorithmicensure{\textbf{Output:}}
\usepackage{tikz}
\usetikzlibrary{shapes.geometric, arrows}
\usepackage{comment}
\usepackage[caption=false]{subfig}

\newcommand{\eg}{{e.g.,}\xspace}

\newcommand{\etal}{{et~al.}\xspace}
\newcommand{\ie}{{i.e.,}\xspace}

\newcommand{\ci}{{\it (i) }}
\newcommand{\cii}{{\it (ii) }}
\newcommand{\ciii}{{\it (iii) }}

\newcommand{\ca}{{\it (a) }}
\newcommand{\cb}{{\it (b) }}
\newcommand{\cc}{{\it (c) }}

\definecolor{notecolor}{rgb}{0.8,0,0} 
\usepackage{amsthm}

\theoremstyle{definition}

\definecolor{scol}{rgb}{0, 0, 1}

\newcommand{\hps}{HPI\xspace}
\newcommand{\has}{HAI\xspace}

\begin{document}

\title{A Simplex-Inspired Architecture for Integrating Quantum Capabilities into Cyber-Physical Systems}


 \author{\IEEEauthorblockN{Tamim Ahmed, Dacheng Shen, Mengyu Liu, and Monowar Hasan}
 \IEEEauthorblockA{School of Electrical Engineering and Computer Science \\
 Washington State University, WA, USA \\
 Emails: tamim.ahmed@wsu.edu, dacheng.shen@wsu.edu, mengyu.liu@wsu.edu, monowar.hasan@wsu.edu}
 }

\maketitle

\begin{abstract}
Cyber-physical systems require accurate and reliable system models to ensure safe and efficient operation. Classical Gaussian Process Regression (GPR) provides uncertainty-aware predictions but suffers from high computational complexity, which limits its scalability in real-time applications. Quantum-assisted Gaussian process models reduce complexity in inference but their practical use is constrained by noise and stability concerns in safety-critical environments. In this paper, we propose a hybrid classical–quantum system identification framework based on a Simplex architecture. The framework combines Quantum-Assisted Hilbert-Space Gaussian Process Regression (QA-HSGPR) as a high-performance module and classical GPR as a high-assurance module. A runtime monitor evaluates system safety and dynamically switches between the two models. Experiments on a Continuous Stirred-Tank Reactor benchmark demonstrate that the proposed framework enables a controllable trade-off between performance and safety for real-time cyber-physical systems.
\end{abstract}
\section{Introduction}
\label{sec:introduction}

Cyber-physical systems (CPS) integrate computational algorithms with physical processes, where an accurate system model directly affects control performance and operational safety. However, in many practical scenarios, the system model is unknown or difficult to derive due to changing operating conditions and uncertainty. Therefore, data-driven system identification methods are proposed to learn the system model directly from data streams.


Gaussian Process Regression (GPR)~\cite{rasmussen2003gaussian} has been widely applied in system identification with uncertainty estimation. However, GPR encounters scalability challenges when applied to high-dimensional data, as kernel evaluation and matrix inversions are computationally expensive. Quantum models can implement feature maps that embed classical data into high-dimensional Hilbert spaces, which enables richer representations and kernel-based learning~\cite{havlivcek2019supervised}. Such quantum-enhanced kernels offer a potential pathway to more efficiently handle high-dimensional nonlinear structures in uncertainty quantification.


Recent studies have begun combining quantum and classical modules, employing quantum kernels to model complex nonlinear dynamics while retaining classical components for interpretability and ease of deployment~\cite{rapp2024quantum}. However, noise and non-determinism in NISQ (Noisy Intermediate-Scale Quantum) devices make it challenging to apply these methods to safety-critical, real-time, closed-loop control~\cite{heyraud2022noisy}.

To address this challenge, we propose a hybrid system identification framework with a Simplex architecture~\cite{seto1998simplex, wang2013l1simplex} for CPS that balances performance and safety. Simplex is a well-known design for CPS in which a \textit{high-performance} unit is primarily responsible for a safety-critical system. This is achieved by the presence of a \textit{high-assurance} component,
and a \textit{decision module} that can take control and maintain the safety of the system if the high-performance module risks pushing the physical plant beyond a precalculated safety envelope.

We employ quantum Gaussian process regression (QGPR)~\cite{farooq2024quantum} as a high-performance identification model to provide stronger nonlinear modeling capabilities, while GPR is used as a high-assurance identification model to maintain more stable and reliable predictions under elevated risk. Additionally, we propose a lightweight real-time monitor that tracks physical safety indicators and their trends, enabling arbitration between the two identification models before the system enters unsafe operating regions. This mechanism allows the quantum model to be applied during stable operation while promptly reverting to the classical model near safety boundaries, thus achieving a controllable trade-off between performance and safety~\cite{phan2017component}.

\section{Background}

\subsection{Classical Gaussian Process Regression}

GPR is a nonparametric method widely used to predict unknown values from known data points~\cite{rasmussen2003gaussian}. It extracts the underlying relations between data points and places a probability distribution directly over functions. We can describe GPR with two quantities: \ci a mean function to represent the expected behavior, and \cii a covariance function also known as the \textit{kernel matrix} to encode the assumptions with variability or smoothness of the function. We will formally define a GPR and discuss its classical limits.

Consider a dataset, $\mathcal{D} = \{\mathbf{x}_i, y_i\}_{i=1}^{N}$, where each $\mathbf{x}_i \in \mathbb{R}^d$ is an input and $y_i$ is the corresponding noisy observation. We model the unknown function as:
\begin{equation}
f \sim \mathcal{GP}(0, k(\mathbf{x}, \mathbf{x}')),\quad
y_i = f(\mathbf{x}_i) + \varepsilon_i,\;
\varepsilon_i \sim \mathcal{N}(0, \sigma_n^2).
\end{equation}

where $k(\mathbf{x}, \mathbf{x}')$ is the kernel matrix, and $\sigma_n^2$ is the variance of the observation noise. The predictive performance is directly related to the choice of its kernel (\ie different kernel functions encode different assumptions). In this work, we are using the squared exponential kernel~\cite{rasmussen2003gaussian}. The kernel is defined as:
\begin{equation}
    k(\mathbf{x}, \mathbf{x}') = \sigma_f^2 \exp\!\left(-\frac{\|\mathbf{x} - \mathbf{x}'\|^2}{2\ell^2}\right),
    \label{eq:se_kernel}
\end{equation}

where $\sigma_f^2$ is a parameter to control the overall variation in amplitude of the function and $\ell$ determines how quickly the function changes with respect to its input (a large $\ell$ means data points remain correlated even when they are far apart, \ie a smooth and slowly varying function). For a new test input $\mathbf{x}_*$, the GP provides a predictive distribution which is: $p(f_* \mid \mathbf{x}_*, \mathcal{D}) = \mathcal{N}(\mu_*, \sigma_*^2)$, where the predictive mean and variance are:
\begin{equation}
\begin{aligned}
    \mu_* &= \mathbf{k}_*^\top \big(\mathbf{K} + \sigma_n^2 \mathbf{I}\big)^{-1} \mathbf{y}, \\
    \sigma_*^2 &= k(\mathbf{x}_*, \mathbf{x}_*) - \mathbf{k}_*^\top \big(\mathbf{K} + \sigma_n^2 \mathbf{I}\big)^{-1} \mathbf{k}_*.
\end{aligned}
\label{eq:gp}
\end{equation}

Here, $\mathbf{y} = [y_1, \dots, y_N]^\top$ is the vector of all observations, $\mathbf{K} \in \mathbb{R}^{N \times N}$ is the kernel matrix where $K_{ij} = k(\mathbf{x}_i, \mathbf{x}_j)$ measures the similarity between training inputs, and $\mathbf{k}_* \in \mathbb{R}^{N}$ is the vector of similarities between the test input and each training point. 

\subsection{Quantum Gaussian Process Regression}
\label{subsec:quantum}
Classical GPR provides a way to make the prediction with uncertainty, but its practical use is limited by the cost of inference. Both predictive mean and variance in Eq.~\eqref{eq:gp} need to solve $(\mathbf{K} + \sigma_n^2 {I})^{-1} \mathbf{y}$. The linear system is typically solved using Cholesky decomposition, which has a time complexity of $O(N^3)$, limiting its scalability for larger datasets. Quantum algorithms offer significant complexity reduction against their classical counterparts~\cite{nielsen2010quantum}. Fidelity-based quantum kernels, instead of using a classical kernel, can preserve the predictive uncertainty~\cite{rapp2024quantum}. Researchers also utilize quantum phase estimation, and HHL~\cite{harrow2004coherent} to propose a coherent-state-based quantum Gaussian process regression algorithm, which can produce quadratic speed-up over classical GPR in inference~\cite{chen2022quantum}. Farooq et al.~\cite{farooq2024quantum} proposed a classical kernel approximation with quantum subroutines that achieves a polynomial speed-up in GPR inference.  

Quantum-Assisted Hilbert-Space Gaussian Process Regression (QA-HSGPR)~\cite{farooq2024quantum} approximates the classical covariance matrix ($\mathbf{K}$) using a Hilbert-space basis expansion. In the approximation, the number of eigenfunctions, $M$, replaces the dependence on the number of training samples $N$, reducing the matrix inversion cost from $O(N^3)$ to $O(M^3)$ in the classical reduced-rank formulation. The accuracy of the approximation depends on the domain parameter, $L$, and the kernel hyperparameters including the amplitude and length scale. $L$ defines the finite spatial domain over which the kernel is approximated using eigenfunctions of the Laplace operator on a bounded interval, $\Omega = [-L, L]$. The number of dominant eigenvalues, $R$, further controls the complexity, which truncates the reduced-rank model and decreases the circuit depth. The quantum circuit implementation requires three main registers: \ci a data register of $\log_2(NM)$ qubits---to encode the Hilbert-space feature matrix in amplitude form, \cii an eigenvalue register with $n_{eig}$ qubits for quantum phase estimation, and \ciii one or two ancilla qubits used to implement controlled rotations, the Hadamard and SWAP tests for computing the predictive mean and variance.  In this work, we are utilizing the QA-HSGPR~\cite{farooq2024quantum} as our predictive framework in our proposed Simplex-based design (see Section~\ref{subsec:simplex}).

\subsection{Simplex Controller Architecture}
Runtime assurance framework---\eg Simplex architecture~\cite{seto1998simplex}---is well-established to guarantee safety in cyber-physical systems. The paradigm of Simplex has three main components: \ca High-Performance Controller (HPC)---provides efficiency but lacks formal guarantees, \cb High-Assurance Controller (HAC)---can be conservative but makes safe operations for system stability, and \cc Decision Logic module---monitors system state continuously and enforces safety by switching control from HPC to HAC for safe operations. Consider a discrete-time cyber-physical system:
\begin{equation}
x_{k+1} = f\big(x_k, u_k\big),
\end{equation}
where $x_k \in \mathbb{R}^n$ denotes the system state at step $k$ and $u_k$ is the control input applied over the sampling interval $[k, {k+1})$. A monitoring function, $g(x_k)$, is the core of the decision logic module that measures the proximity of the system to a predefined safe set $\mathcal{S}$. The decision module determines the active controller mode and applies the control input.

\begin{figure}[t]
    \centering
    \includegraphics[width=1\linewidth]{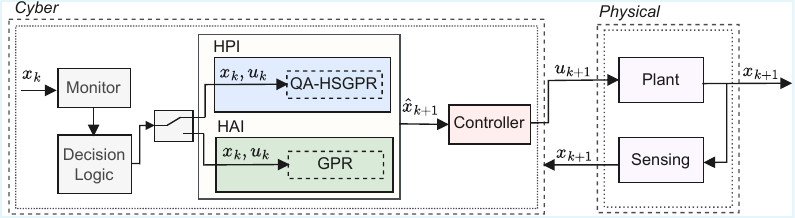}
    \caption{Overview of the proposed Simplex-based hybrid system identification architecture integrating a high-performance identification (\hps) and a high-assurance identification (\has) with a decision logic module for safety-critical cyber–physical systems.} 
    \label{fig:simplex}
\end{figure}

In this work, we want to propose a new Simplex-motivated architecture (see Fig.~\ref{fig:simplex}) that consists of: \ca \textit{High-Performance Identification (\hps)}---a quantum assisted Gaussian process regression that is computationally efficient and scalable, and \cb \textit{High-Assurance Identification (\has)}---a classical Gaussian process regression which provides highly accurate and reliable predictions. The decision module continuously monitors the system states and selects the appropriate identification module (\hps or \has). Based on the decision logic, the controller then subsequently generates the control input for the plant accordingly.

\subsection{CSTR Model}
\label{subsec:cstr}

Our framework is evaluated on the Continuous Stirred-Tank Reactor (CSTR) benchmark~\cite{seborg2016process}, an irreversible exothermic reaction process conducted in a vessel. The system state is determined by the reactant concentration $C_A$ and the reactor temperature $T$. The control input $u$ represents the temperature of the cooling jacket, which removes the heat generated by the reaction through heat exchange, thereby regulating the system temperature. The core challenge of this system lies in its strongly nonlinear kinetics. Because the reaction rate is highly sensitive to temperature, even a small increase in temperature can lead to a sharp acceleration of the reaction rate, resulting in additional heat release. This positive feedback loop, where a temperature increase accelerates the reaction and heat generation, which further increases temperature, can lead to thermal runaway. Therefore, under disturbances, if the cooling capacity is insufficient to offset the generated heat, the system can approach and potentially violate the safety boundary. This inherent risk motivates the introduction of the Simplex architecture for real-time safety monitoring in this paper. The simulation model and parameters used in this study are adopted from prior work~\cite{seborg2016process}.


\label{sec:system_model}
\label{subsec:simplex}

\section{Overview of the Proposed Framework}
Our framework adapts from a single-layer switching Simplex architecture~\cite{sha2001using}, consisting of HPI, HAI, and a monitor. Different from the original simplex architecture, which arbitrates between two controllers, the proposed framework applies runtime assurance at the system identification level, where a single shared controller operates based on models provided by two identification modules. As shown in Fig.~\ref{fig:simplex}, when the safety margin is sufficient, the monitor prioritizes the HPI module for higher-performance prediction. When the system approaches a safety boundary or an increasing risk trend is detected, the monitor triggers a switch to the HAI module to ensure stable and reliable system behavior. The modules in the diagram are: \ci Monitor Module, \cii HPI Module, and \ciii HAI Module. These components are briefly described below.

\noindent \textbf{\textit{\ul{Monitor Module:}}}
At each time step $k$, the monitor maps the measured states to a dimensionless safety metric $\Sigma_k$. The switching logic compares $(\Sigma_k, s_k)$ with $\Sigma_{\mathrm{warn}}$ and $\Sigma_{\mathrm{crit}}$ to determine which identification model will be used for the next prediction.

\noindent \textbf{\textit{\ul{HPI Module:}}} The HPI module takes the current state measurements as input and, under normal operating conditions, outputs a high-performance next-step state prediction.

\noindent \textbf{\textit{\ul{HAI Module:}}} The HAI module takes the same input as the HPI module and outputs a more reliable next-step state prediction near the safety boundary.





\begin{table}[t]
\footnotesize
\centering
\caption{Nomenclature and symbols used in this article}
\label{tab:switching_nomenclature}
\begin{tabular}{|c|l|}
\hline
Symbol & Description \\
\hline\hline
$T_k$ & Measured reactor temperature at time step $k$ \\
$C_{A,k}$ & Measured reactor concentration at time step $k$ \\
$T_{\text{mean}}$ & Mean temperature \\
$C_{A,\text{mean}}$ & Mean concentration \\
$T_{\text{limit}}$ & Temperature safety boundary \\
$C_{A,\text{limit}}$ & Concentration safety boundary \\
$\Delta_T(k)$ & Normalized temperature deviation at step $k$ \\
$\Delta_C(k)$ & Normalized concentration deviation at step $k$ \\
$\Sigma_k$ & Composite safety index at step $k$ \\
$s_k$ & Rate of change of $\Sigma_k$ \\
$\Sigma_{\text{crit}}$ & Critical threshold \\
$\Sigma_{\text{warn}}$ & Warning threshold \\
$m_k$ & Applied control input at time step $k$ \\ 
\hline
\end{tabular}
\end{table}


\section{Online Switching Logic}
\label{sec:switching}

This section specifies the online switching rule used by the monitor to switch between the two identification methods. The core idea behind the switching logic is that when we are close to the safety boundary, we switch to the HAI for assurance of safety. Otherwise, we apply the HPI for better performance. All the notations used for CSTR modeling and switching conditions are summarized in Table~\ref{tab:switching_nomenclature}.

\subsection{Safety Index Construction}
\label{subsec:safety_index}
Following the foundational simplex runtime-assurance principle~\cite{seto1998simplex} and its component-based extensions~\cite{phan2017component}, we define a physics-based safety index, $\Sigma_k$ to quantify proximity to the safety boundary. Safety boundaries $T_{\mathrm{limit}}$ and $C_{A,\mathrm{limit}}$ are defined on each dimension with different scales. Therefore, we define the normalized deviations from the nominal operating point $(T_{\mathrm{mean}}, C_{A,\mathrm{mean}})$:
\begin{equation}
\label{eq:normalized_deviation}
\Delta_T(k)=\frac{T_k-T_{\mathrm{mean}}}{T_{\mathrm{limit}}}, \qquad
\Delta_C(k)=\frac{C_{A,k}-C_{A,\mathrm{mean}}}{C_{A,\mathrm{limit}}}.
\end{equation}
The composite safety index is then defined as
\begin{equation}
\label{eq:sigma_def}
\Sigma_k=\sqrt{\Delta_T(k)^2+\Delta_C(k)^2}.
\end{equation}

Based on $T_{\mathrm{limit}}$ and $C_{A, \mathrm{limit}}$, we construct a safety metric, $\Sigma_k$ using the Euclidean norm to aggregate risks from multiple variables. A system is considered safe if it remains within the half-space defined by $\Sigma_k < 1$, where $\Sigma_k = 1$ represents the one-sided boundary. This formulation explicitly accounts for aggregate risk---even if individual variables remain within their respective independent half-space constraints, their simultaneous proximity to the limits can cause the composite metric to breach the half-space boundary, thereby triggering a safety violation. To capture the risk trend, we compute the first-order deviation of the safety index, $s_k=\Sigma_k-\Sigma_{k-1}$, where $s_k > 0$ indicates an increasing risk, while $s_k \le 0$ indicates a stable or recovering trend.


\subsection{Switching Rule}
\label{subsec:switch_rule}

At each time step $k$, the monitor outputs the active mode $m_k \in \{\text{HPI},\text{HAI}\}$ based on the safety metrics $(\Sigma_k, s_k)$. The switching policy is defined as:
\begin{equation}
\label{eq:switching_rule}
m_k=
\begin{cases}
\text{\has}, & \text{if } \Sigma_k \ge \Sigma_{\mathrm{crit}}
\ \vee\ \big(\Sigma_k \ge \Sigma_{\mathrm{warn}} \wedge s_k > 0\big), \\[6pt]
\text{\hps}, & \text{otherwise}.
\end{cases}
\end{equation}

where $\Sigma_{\mathrm{crit}}$ is the critical threshold and $\Sigma_{\mathrm{warn}}$ is the warning threshold, satisfying $0 < \Sigma_{\mathrm{warn}} < \Sigma_{\mathrm{crit}}$. This two-threshold design adopts the "\textit{Operational Envelope}" and "\textit{Safety Envelope}" concept from the original Simplex architecture~\cite{sha2001using}. There are three conditions to switch back and forth between \has and \hps:

\subsubsection{Reactive fallback ($\Sigma_k \ge \Sigma_{\mathrm{crit}}$)} The system reaches or exceeds the critical boundary, indicating a higher safety risk. 


\subsubsection{Proactive fallback ($\Sigma_k \ge \Sigma_{\mathrm{warn}}$ and $s_k>0$)} The system enters a warning region while the risk continues to increase. The framework triggers a proactive fallback to reduce the likelihood of crossing the boundary at the next step. 


\subsubsection{Normal operation (otherwise)} The system has sufficient safety margin, or it is near the boundary, but the risk is not increasing ($s_k\le 0$). The framework keeps HPI active to benefit from higher predictive performance. 


We provide a concrete example of the switching logic. Consider $\Sigma_{\mathrm{crit}}=1$ as the hard safety limit. The warning threshold is set to $\Sigma_{\mathrm{warn}}=0.8$, creating a $20\%$ safety buffer to compensate for the system's thermal inertia and discrete sampling delays, ensuring the HAI has sufficient time to intervene before the critical boundary is breached. When $T_k$ and $C_{A,k}$ are close to the safety boundary (but do not exceed), the safety index remains within the warning region ($0.8 < \Sigma_k < 1$). Although within the physical boundary ($\Sigma_k < 1$), the fact that $s_k > 0$ (indicating $\Sigma_k > \Sigma_{k-1}$) can still trigger a preventive fallback to HAI mode to ensure the system is working under a stable region. Conversely, in the recovery phase, the safety index may remain within the warning region ($0.8 < \Sigma_k < 1$). However, the negative slope ($s_k < 0$ and $\Sigma_k < \Sigma_{k-1}$) indicates that the system is stabilizing, allowing a return to HPI mode. Finally, if the deviation exceeds the safety boundary \ie when $|T_k - T_{\mathrm{mean}}| > T_{\mathrm{limit}}$ or $\Sigma_k \ge 1$, a hard fallback will be unconditionally triggered, overriding any trend analysis.

\section{Evaluation}
We evaluated the predictive performance of classical GPR and QA-HSGPR to highlight the trade-off between accuracy and complexity. Additionally, we presented the runtime switching behavior of the proposed Simplex framework to demonstrate the safety-critical operating conditions.

\subsection{Experimental Setup}
\subsubsection{CSTR Data Generation}
We use a high-fidelity simulation environment, cpsim~\cite{liu2024cpsim}, to generate an appropriate dataset for our CSTR model. The data points were generated with a sampling interval of 0.02~s under a piecewise-constant temperature reference signal that varies within the range of 290-310~K to excite nonlinear operating conditions. We recorded the reactor concentration ($C_{A}$, $C_{A,ref}$) and temperature ($T_k$, $T_{ref}$) with the control input at each time step. 64 uniformly sampled data points have been selected in a time window of $[25, 95]$~s, where 32 of them are used for training, and the rest are used as testing data points. 

\subsubsection{Data Preprocessing}
\label{subsec:dp}

We transformed the raw CSTR simulated data into normalized and reduced-dimensional inputs for regression analysis. We extracted three physical state variables-- \ci reactant concentration ($C_{A,k}$), temperature ($T_{k}$), and control input ($u_k$) for each time step. The input feature is ${x}_k = [C_{A,k}, T_{k}, u_k]^\top$ and the output feature is $\hat{x}_{k+1} =[T_{k+1}]$. We performed $\sigma_{\text{noise}} = \text{std}(\Delta T)/\sqrt{2}; \;\forall \Delta T$, where $\Delta T= T_{k+1}-T_{k}$. Both input and output feature are normalized using StandardScaler~\cite{pedregosa2011scikit}. The three-dimensional input feature was further reduced to one dimension using Principal Component Analysis (PCA)~\cite{abdi2010principal}. In our case, it retains 91.65\% of the total variance.

\subsubsection{GP-Specific Configuration}
For both classical GPR and QA-HSGPR, we used the same kernel (squared exponential kernel). Two kernel parameters, $\sigma_f=1$ and $\ell=1.5$, are chosen for Eq.~\eqref{eq:se_kernel} to control variance and correlations of the GP output. We calculated the likelihood variance, $\sigma^2$, from $\sigma_{\text{noise}}$ and normalized the standard deviation. For QA-HSGPR, we need to transform the kernel from the $N$-dimensional Gaussian process into an $M$-dimensional approximation. We use $M=2^4=16$ eigenfunctions on the bounded domain $[-L,L]$, where $L= 3*|\mathbf{X}_{\text{PCA}}|$ for Hilbert-Space Approximation. The quantum algorithm uses Quantum Phase Estimation (QPE) to accelerate computations on the $M \times M$ feature matrix. The quantum parameters used in our analysis are: \ci $n_{\text{eig}} = 13$ qubits to determine how accurately eigenvalues are measured, \cii $N_{\text{shots}} = 10^6$ repeated measurements to reduce statistical noise in quantum results, and \ciii rank, $R = 4$, which keeps only the top four most important components and truncates the rest to filter out quantum noise. Additionally, a small regularization shift $\delta = \lambda_{\max} + 0.012$ ($\lambda_{\max}$ is the largest eigenvalue of the normalized feature covariance matrix) was added to prevent numerical instability during matrix inversion.

\subsection{Prediction Analysis: Classical GPR vs QA-HSGPR}

\begin{figure}[t]
    \centering \includegraphics[width=1.0\linewidth]{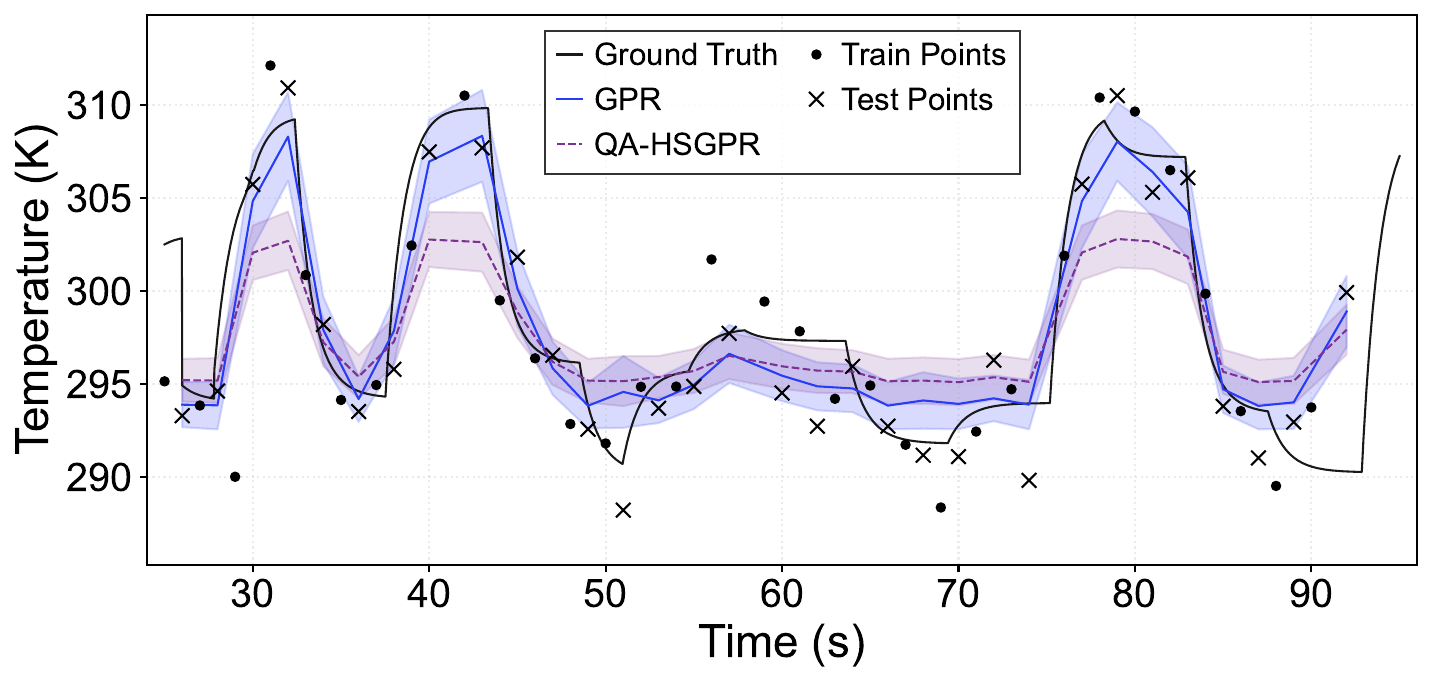}
    \caption{Temperature ($T_k$) prediction comparison on the CSTR dataset. The ground-truth trajectory is shown along with training and testing samples. Classical GPR provides closer tracking of the nonlinear dynamics and sharper uncertainty bounds, while QA-HSGPR demonstrates slightly lower accuracy.}
    \label{fig:prediction}
\end{figure}

Fig.~\ref{fig:prediction} illustrates the temperature prediction result for both Classical GPR and QA-HSGPR on the CSTR dataset. We can observe that the classical method captures nonlinear temperature variations better than approximation-based quantum prediction, especially in regions with rapid changes in the system dynamics. Similarly, Table~\ref{tab:prediction_metrics} summarizes the quantitative analysis for predicting the test points using MAE, MSE, and RMSE. The results demonstrate that classical GPR produces lower error while predicting across all metrics. 

Recall that \has needs accuracy to produce safe behavior for system stability. As classical GPR can provide results closer to the ground truth, it can assure safety when the system has passed the safest border of stability. But the bottleneck issue described in Section~\ref{subsec:quantum} limits the scalability of Classical GPR. Similarly, \hps prefers efficiency over accuracy. Because of reduced-rank based approximation, QA-HSGPR can provide an reduce the time complexity and asymptotic speed-up compared to classical GPR. Therefore, QA-HSGPR has been chosen as the integrated predictive framework for \hps.
 
\begin{table}[t]
\centering
\caption{Prediction performance comparison on the CSTR dataset.}
\label{tab:prediction_metrics}
\renewcommand{\arraystretch}{1.4}
\begin{tabular}{l|c|c|cl}
\hline
Model & MAE & MSE & RMSE \\
\hline
\hline
Classical GPR & 2.453 & 10.127 & 3.182 \\
QA-HSGPR      & 2.882 & 12.990 & 3.604 \\
\hline
\end{tabular}
\end{table}

\subsection{Runtime Switching Behavior and Safety Analysis}

\begin{figure}[t]
    \centering \includegraphics[width=1.0\linewidth]{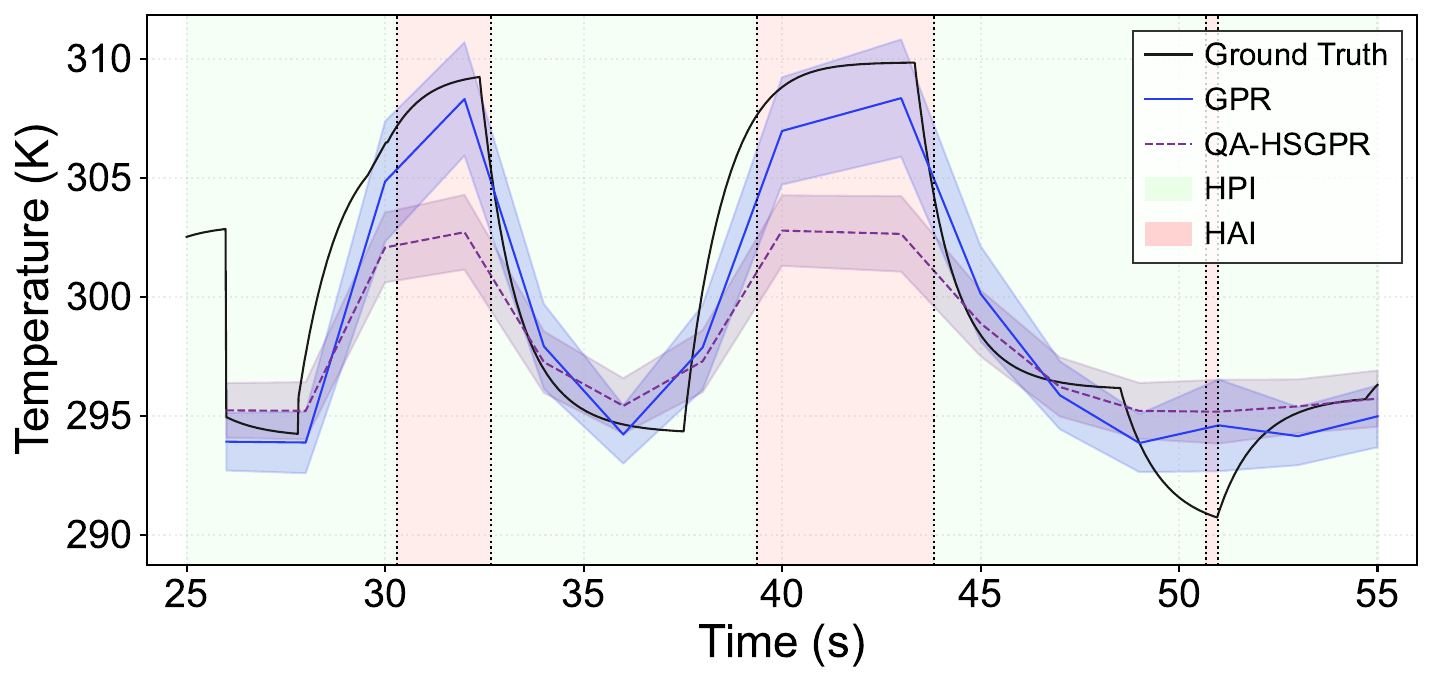}
    \caption{Runtime switching of the proposed simplex-based hybrid framework in the CSTR process. Transitions from the high-performance (\hps) and high-assurance (\has) modules are triggered by the safety index and its trend. Green regions correspond to predictions from \hps, whereas red regions correspond to \has.}
    \label{fig:switch}
\end{figure}

In Fig.~\ref{fig:switch}, we illustrate the online switching behavior of the proposed simplex-based hybrid framework for the time window $[25,55]$~s. The figure highlights the transitions between \hps (green region) to \has (red region) based on the safety index $\Sigma_k$ and $s_k$. The warning region ($0.8<\Sigma_k<1$) evaluates both the magnitude and the direction of the safety index. A positive slope, $s_k>0$, indicates that the system is drifting toward the safety boundary, and a negative slope $s_k<0$ reflects the nominal operating region. For example, at $t=30.3$~s, the safety index, $\Sigma_k=0.8050$, and slope, $s_k>0$, indicate an increasing safety risk. Therefore, the controller switches from \hps to \has for stabilizing the system before the critical boundary. Similarly, at $t=32.66$~s, the safety index decreases within the warning region, which shows a recovery trend and the system resumes \hps operation while maintaining safety.


\section{Related Work}
\label{sec:related_work}

Kocijan et al.~\cite{kocijan2005dynamic} pioneered the application of GPR for nonlinear dynamic systems, demonstrating its distinct advantages in handling small-sample data. However, standard GPR algorithms are limited by a computational complexity of $\mathcal{O}(N^3)$~\cite{rasmussen2003gaussian}, which restricts their scalability in high-dimensional feature mappings or complex tasks requiring real-time responsiveness.

Quantum Gaussian processes have emerged as a hybrid quantum–classical extension of kernel-based Bayesian learning, where a quantum feature map induces the covariance function of a Gaussian process, building on the kernel methods introduced by Havlicek et al.~\cite{havlivcek2019supervised}, Chen et al.~\cite{chen2022quantum} and Farooq et al.~\cite{farooq2024quantum} developed QGPR variants to capture correlations that are challenging for classical kernels. Despite these theoretical advancements, the inherent stochasticity of NISQ devices remains a barrier to direct deployment, necessitating runtime safeguards~\cite{nieman2024safety}.

Lui \etal~\cite{sha2001using} introduced the Simplex architecture in his seminal paper, which uses decision logic to arbitrate between high-performance and safety-guaranteed modules when the system approaches safety boundaries. Phan et al.~\cite{phan2017component} further extended this framework for systems containing complex, uncertified components. While Simplex is well established for classical black-box models, integrating QGPR into a Simplex runtime assurance framework remains unexplored. This work addresses this gap by leveraging Simplex to manage quantum noise while utilizing the computational potential of quantum kernels. 

\section{Conclusion}

This paper presents a hybrid classical-quantum system identification framework for safety-critical CPS. The proposed approach balances computational efficiency and prediction reliability with its runtime switching behavior. The reduced computational complexity offered by quantum computing makes the high-performance module robust and efficient. On the other hand, the classical algorithm can achieve higher predictive accuracy, and its integration with a high-assurance module can enable reliable decisions for the safest operation. By integrating quantum speed-up with classical computing, this work demonstrates a practical pathway for deploying quantum-classical hybrid framework in real-time control systems.

\balance
\bibliographystyle{ieeetr}
\bibliography{references/reference}


\end{document}